\documentstyle[aps]{revtex}
\begin{document}
\voffset=0mm
%
\title{Superconductivity in the Attractive Hubbard Model: The 
Double Hubbard--I Approximation.}
\author{{\it J.J. Rodr\'{\i}guez-N\'u\~nez}}
\address{Departamento de F\'{\i}sica, Universidade Federal de Santa 
Maria, 97105-900 Santa Maria/RS  
Brazil.\\ e-mail: jjrn@ccne.ufsm.br}
\author{H. Ghosh}
\address{Instituto de F\'{\i}sica, Universidade Federal Fluminense, 
Av. Litor\^anea S/N, Boa Viagem, 24210-340 Niter\'oi/RJ, Brazil. 
e-mail: hng@if.uff.br}
\date{\today}
\maketitle

%
%
\begin{abstract}
Using the Dyson equation of 
motion for both the diagonal one-particle 
Green function, $G(\vec{k},\omega)$ 
and off--diagonal Green function, $F(\vec{k},\omega)$, 
at the level of the Hubbard--I decoupling scheme, we have 
found that they have four poles symmetric in pairs, 
justifying a more elaborated calculation done by the 
Z\"urich group by means of the $T$-Matrix approach 
(Pedersen et al, Z. Physik B {\bf 103}, 21 (1997)) and 
the moment approach of  Nolting 
(Z. Physik {\bf 255}, 25 (1972)).  
We find that the energy spectra and the 
weights of $G(\vec{k},\omega)$ and 
$F(\vec{k},\omega)$ have to be calculated 
self-consistently. $G(\vec{k},\omega)$ satisfies 
the first two moments while  
$F(\vec{k},\omega)$ the first sum rule. 
Our {\it order parameter} $\alpha(T)$ is
given by $1/N_s \sum_{{\vec{k}}}
\varepsilon({\vec{k}})\Delta({\vec{k}})$. 
Due to the fact that we have 
a purely local attractive interaction $\Delta(\vec{k})$ can be 
of {\it any} s--type wave. However, for 
a {\it pure s--wave}, for which $\alpha(T) = 0$, we go back to the
mean--field $BCS$ results, with a renormalized chemical potential.
In this case, the off--diagonal Green function, $F(\vec{k},\omega)$,
satisfies the first two off--diagonal sum rules.  
We explicitly state the range of validity of our approximation.\\
\\
Pacs numbers: 74.20.-Fg, 74.10.-z, 74.60.-w, 74.72.-h
\end{abstract}

%
%
%
%

	After the discovery of 
the high-$T_c$ materials
\cite{Bednorz-Muller}, the study of correlations 
has gained interested due to the fact that there is 
the belief\cite{Anderson} 
that the normal properties of these materials 
could be explained in the framework of the Hubbard 
model\cite{HubbardI,HubbardII}, since electron correlations 
are strong, i.e., the on-site electron-electron 
repulsions $U$ are much larger than the energies 
associated with the hybridization of atomic orbitals 
belonging to different atoms\cite{Fulde}. This Hamiltonian is a kind 
of minimum model\cite{tasaki} which takes into account quantum 
mechanical motion of electrons in a solid, and nonlinear repulsion 
between electrons. Even though this model is too simple to 
describe solids faithfully, serious theoretical studies have 
revealed that to understand the various properties of it is a 
very difficult task. Its study will prove useful in developing 
various notions and techniques in statistical physics of 
many particle physics.

	Since the high-$T_c$ superconducting materials are 
extreme type II superconductors, with a short coherence 
volume, one might take this fact as an indication of tightly 
bound pairs or/and that correlations effects 
strongly affect the properties of such materials. However, 
this scenario has been challanged by recent tunneling 
measurements\cite{miyakawa} which lead to the conclusion 
that, for example, underdoped $Ba_2Sr_2CaCu_2O_{8-\delta}$ 
is described by intermediate coupling interaction, 
because the pairing fluctuations persist up to 
$T^*$\cite{RandVar}, where $T^*$ ($T^* > T_c$) is the temperature of 
pair fluctuations and $T_c$ is the superconducting critical 
temperature. 

	One of the 
simplest model featuring superconductivity and allowing a 
systematic study of correlations is the attractive Hubbard
model\cite{Micnas}  
which we adopt in this paper as the counterpart of the usual 
Hubbard model. This model has been used to 
explore qualitative features of the superconducting phase 
transition\cite{Randeria}. Ref.\cite{Randeria} is mainly a  
review of the analytical work done on this model.  
Recently, Huscroft and Scalettar\cite{HS} find that the superconducting 
order parameter is more stable that the charge density wave order 
at half-filling in the presence of disorder. Then, due to these 
considerations, we concentrate in the superconducting properties 
leaving outside any treatment of charge density 
wave order.

	We will use the Dyson equation of 
motion technique\cite{FW} for both the diagonal,  
$G(\vec{k},\omega)$, and 
off-diagonal one-particle Green function,   
$F(\vec{k},\omega)$, at the level of Hubbard--I 
decoupling scheme\cite{HubbardII}. The main theoretical conclusion 
of this paper is that both the diagonal and off 
diagonal one-particle Green functions have four poles. 
These poles are symmetric in pairs 
verifying a more elaborated calculation of the 
Z\"urich group\cite{1}.  

%

    	The model we study is the  
Hubbard model\cite{Micnas}
\begin{eqnarray}\label{Ham}
H = t_{\vec{i},\vec{j}}c_{\vec{i}\sigma}^{\dagger}c_{\vec{j}\sigma}
   + \frac{U}{2} n_{\vec{i}\sigma}n_{\vec{i}\bar{\sigma}}   
   - \mu c^{\dagger}_{\vec{i}\sigma}c_{\vec{i}\sigma}~~,
\end{eqnarray}
where $c_{\vec{i}\sigma}^{\dagger}$($c_{\vec{i}\sigma}$) are creation
(annihilation) electron operators with spin $\sigma$. $n_{\vec{i}
\sigma} \equiv c_{\vec{i}\sigma}^{\dagger}c_{\vec{i}\sigma}$. 
$U = - |U|$ is the local attractive interaction and $\mu$ the chemical
potential (we work in the grand canonical ensemble). We have 
adopted Einstein convention for repeated 
indices, i.e., for the $N_s$ sites $\vec{i}$, the 
$z$ nearest-neighbor sites 
$\vec{j}$ and for spin up and down ($\sigma = -\bar{\sigma} 
= \pm 1$). $t_{\vec{i},\vec{j}} = -t$, for n.n. 
and zero otherwise. Other types of hopping, i.e., 
$t' \neq 0$ between next nearest neighboors $(n.n.n.)$ 
could be considered in our formalism\cite{Roland}. In 
this paper, we restrict ourselves to $n.n.$ 
hoping.

	We need to evaluate the equation of motion\cite{FW} 
for the operators 
$c_{\vec{i}\sigma}$ and $c^\dagger_{\vec{i}\sigma}$. They are

\begin{equation}\label{dyn}
\frac{\partial  c_{\vec{i}\sigma}}{\partial \tau} = 
\left[ c_{\vec{i}\sigma}, H \right]_{\bf -} 
~~~ ; ~~~\frac{\partial c^\dagger_{\vec{i} \sigma}}{\partial \tau}  =  
\left[ c^\dagger_{\vec{i} \sigma}, H \right]_{\bf -}~~~,
\end{equation}

\noindent
where the sign $\bf -$ means the commutator. 
Combining Eqs. (\ref{Ham},\ref{dyn}) we get

\begin{eqnarray}\label{needdyn}
\frac{\partial c_{\vec{i} \sigma}}{\partial \tau}  
= + t_{\vec{i},\vec{l}} c_{\vec{l} \sigma} 
+ U c_{\vec{i} \sigma} n_{\vec{i} \bar \sigma} ~~~; ~~~
 \frac{\partial  c^\dagger_{\vec{i} \sigma}}{\partial \tau} =  
- t_{\vec{i},\vec{l}} c^\dagger_{\vec{l} \sigma} 
- U c^\dagger_{\vec{i} \sigma} n_{\vec{i} \bar \sigma} ~~~.
\end{eqnarray}

	Next, the one-particle Green's function is 
defined as
\begin{equation}\label{1PGF}
G_{\sigma}(\vec{i},\vec{j};\tau) \equiv - 
\left< \left< T_{\tau}c_{\vec{i}\sigma}(\tau);
c^{\dagger}_{\vec{j}\sigma}(0) 
\right> \right>~~~,
\end{equation}
\noindent where $T_{\tau}$ means time ordering. Combining 
Eqs. (\ref{needdyn},\ref{1PGF}),  and 
Fourier analyzing the time and space 
variables we end up with the following 
equation for $G(\vec{k},\omega)$

\begin{equation}\label{DEM} 
(\omega - \varepsilon_{\vec{k}})G(\vec{k},\omega) = 
1 + U \Gamma^{(2)}(\vec{k},\omega)~~~,
\end{equation}
\noindent where $\Gamma^{(2)}(\vec{k},\omega)$ is the 
Fourier transform of the doubly occupied Green 
function\cite{EA}

\begin{equation}\label{Gamma2}
\Gamma^{(2)}(\vec{k},\omega) \equiv \left< 
\left< n_{\vec{i}\bar{\sigma}}(\tau) 
c_{\vec{i}\sigma}(\tau); c^{\dagger}_{\vec{j}\sigma}(0) 
\right> \right> (\vec{k},\omega)~~~.
\end{equation}

	As we see from Eq. (\ref{Gamma2}), the one--particle 
Green function and the doubly occupied Green function are 
connected thruout the equation of motion. We get the 
doubly occupied Green function mainly due to the presence 
of four operators in the Hubbard interaction. In Eq. (\ref{Gamma2}), 
$n_{i\bar{\sigma}} \equiv c^{\dagger}_{i\bar{\sigma}}c_{i\bar{\sigma}}$ 
is the occupation number operator. Our next step  
is to apply the Dyson equation to $\Gamma^{(2)}(\vec{k},\omega)$, 
which is given by 
\begin{equation}\label{DysonGamma}
\left(\omega - U \right) \Gamma^{(2)}(\vec{k},\omega) = 
\rho_{\bar{\sigma}} + t_{{\vec{i}},{\vec{l}}} \left< \left< 
 n_{\vec{i}\bar{\sigma}}(\tau)
c_{\vec{l}\sigma}(\tau); c^{\dagger}_{\vec{j}\sigma}(0)
\right> \right> ({\vec{k}},\omega)~~~,
\end{equation}
\noindent which we have obtained under the assumption that 
\begin{equation}\label{approximation}
\frac{\partial n_{\vec{i},\sigma}}{\partial \tau} = 0~~~.
\end{equation}

	Eq. (\ref{approximation}) is certainly an 
approximation which reproduces the Hubbard--I solution 
in the equation of motion approach and it is {\it only}  
valid at the level of Eq. (\ref{DysonGamma}). We leave for the 
future\cite{JJRNpre} the study of the effect of 
going beyond the approximation given by Eq. (\ref{approximation}). 
Now we perform a decoupling in the spirit of 
Hubbard\cite{HubbardII} as follows
\begin{equation}\label{decoupling}
\left< \left<
 n_{\vec{i}\bar{\sigma}}(\tau)
c_{\vec{l}\sigma}(\tau); c^{\dagger}_{\vec{j}\sigma}(0)
\right> \right> \approx 
\rho_{\bar{\sigma}} G_{{\vec{l}},{\vec{j}}}(\tau) - 
\frac{1}{U} 
\Delta_{{\vec{l}},{\vec{i}}}F^\dagger_{{\vec{i}},{\vec{j}}}~~~. 
\end{equation}

	Combining Eqs. (\ref{DysonGamma},\ref{decoupling}), we 
arrive to the following expression for $\Gamma_2(\vec{k},\omega)$ 
\begin{equation}\label{123123}
\left( \omega - U \right) \Gamma^{(2)}(\vec{k},\omega) =
\rho_{\bar{\sigma}} + \rho_{\bar{\sigma}} \varepsilon_{{\vec{k}}} 
G(\vec{k},\omega) - \frac{\alpha(T)}{U}F^\dagger (\vec{k},\omega) 
~~~,
\end{equation} 
\noindent where $\alpha(T) = 1/N_s\sum_{{\vec{k}}}\varepsilon({\vec{k}})
\Delta({\vec{k}})$. Combining Eqs. 
(\ref{DEM},\ref{DysonGamma},\ref{123123}) we get
\begin{equation}\label{offcoupling}
\left[(\omega - \varepsilon_{{\vec{k}}})(\omega - U) - 
\rho_{\bar{\sigma}} U \varepsilon_{{\vec{k}}} \right] 
G(\vec{k},\omega) = \omega - U(1 - \rho_{\bar{\sigma}}) - 
\alpha(T)F^\dagger (\vec{k},\omega) ~~~.
\end{equation}

	Thus, from Eqs. (\ref{offcoupling}), we see that 
$G$ and $F$ are coupled and that $G$ reduces to the 
Hubbard--I solution, as it should be, when 
$\alpha(T) = 0$. Let us pause for a while to explain the 
notation. The parameter $\alpha(T)$ is going to be our 
order parameter, in complete analogy with the decoupling 
scheme in mean--field treatments ($BCS$ one, for 
example). Next, we have to find the time evolution for 
$F^\dagger (\vec{k},\omega)$. A similar analysis, i.e., 
another Hubbard--I decoupling scheme for 
$F^{\dagger}(\vec{k},\omega)$, shows that

\begin{equation}\label{offcoupling2}
\left[(\omega + \varepsilon_{{\vec{k}}})(\omega + U) - 
\rho_{\bar{\sigma}} U \varepsilon_{{\vec{k}}} \right] 
F^\dagger(\vec{k},\omega) = \alpha^*(T) G(\vec{k},\omega) 
\end{equation}

	Eqs. (\ref{offcoupling},\ref{offcoupling2}) produce for 
$G(\vec{k},\omega)$ and $F^\dagger(\vec{k},\omega)$ the 
following solutions,
\begin{eqnarray}\label{solutions}
G(\vec{k},\omega) &=& \frac{\left[ \omega - U(1- 
\rho_{\bar{\sigma}}) \right] 
\left[(\omega + \varepsilon_{{\vec{k}}})(\omega + U) - 
\rho_{\bar{\sigma}} U \varepsilon_{{\vec{k}}} \right]} {
\left[(\omega - \varepsilon_{{\vec{k}}})(\omega - U) - 
\rho_{\bar{\sigma}} U \varepsilon_{{\vec{k}}} \right] 
\left[(\omega + \varepsilon_{{\vec{k}}})(\omega + U) - 
\rho_{\bar{\sigma}} U \varepsilon_{{\vec{k}}} \right] + 
\mid \alpha(T) \mid^2} ~~~,\nonumber \\
F(\vec{k},\omega) &=& \frac{\alpha(T) \left[ \omega - U(1- 
\rho_{\bar{\sigma}}) \right]} {
\left[(\omega - \varepsilon_{{\vec{k}}})(\omega - U) -  
\rho_{\bar{\sigma}} U \varepsilon_{{\vec{k}}} \right]   
\left[(\omega + \varepsilon_{{\vec{k}}})(\omega + U) -  
\rho_{\bar{\sigma}} U \varepsilon_{{\vec{k}}} \right] +         
\mid \alpha(T) \mid^2} ~~~.
\end{eqnarray}	

	From Eqs. (\ref{solutions}) we conclude that, 
for $\alpha(T) \neq 0$, 
\begin{eqnarray}\label{poleslike}
G(\vec{k},\omega) = \sum_{j=1}^4 \frac{\hat{\alpha}_j
(\vec{k})}{\omega - \hat{\Omega}_j(\vec{k})}~~~,~~~
F(\vec{k},\omega) = \sum_{j=1}^4 \frac{\hat{\beta}_j 
(\vec{k})}{\omega - \hat{\Omega}_j(\vec{k})}~~~,
\end{eqnarray}
\noindent i.e., the one-particle Green functions have four 
poles (lifetime effects are neglected here) which turn out to be

\begin{eqnarray}\label{fourpoles}
\hat{\Omega}_1(\vec{k}) = - ~\hat{\Omega}_2(\vec{k}) = \omega_o
(\vec{k})~~~,~~~
\hat{\Omega}_3(\vec{k}) = - ~\hat{\Omega}_4(\vec{k}) = \omega_1
(\vec{k})~~~,
\end{eqnarray}
\noindent where
\begin{eqnarray}\label{polesolution}
\omega^2_{o,1}(\vec{k}) &\equiv& 
\frac{1}{2} \left[C(\vec{k}) \pm 
\left[ C^2(\vec{k}) - 4\left(\mid \alpha(T) \mid^2 + 
(1- \rho_{\bar{\sigma}})^2U^2 \varepsilon^2_{{\vec{k}}}\right) 
\right]^{1/2} \right]~~~,\nonumber\\
C(\vec{k}) &\equiv& 
U^2+\varepsilon^2_{\vec{k}}+2U\varepsilon_{\vec{k}}~~~;
~~~\varepsilon_{{\vec{k}}} 
\equiv \varepsilon({\vec{k}}) - \mu~~~,
\end{eqnarray} 
\noindent with $\varepsilon({\vec{k}}) = -2t \sum_{j=1}^d \cos(k_j)$ 
and $d$ the lattice dimension. Taking a closer look to 
$\omega^2_{o,1}(\vec{k})$ in 
Eqs. (\ref{polesolution}) we conclude that these poles have almost 
the form of the poles for $G(\vec{k},\omega)$ and $F(\vec{k},\omega)$ 
obtained in Ref.\cite{1}, since they give four solutions, symmetric 
in pairs, respecting what we call the {\it BCS symmetry}. In 
Ref.\cite{1}, we interpreted these two symmetric solutions as 
corresponding to the $BCS$ solution (the opening of the $BCS$ 
gap around the chemical potential) and to the {\it pair physics} 
(the correlation gap), respectively. In other words, 
our two Hubbard--I decouplings have given an additional contribution, 
which is due to the presence of $pair$ $fluctuations$ 
above $T_c$ and which remain for $T < T_c$. These $pair$ 
$fluctuations$ are the electrons which are not in the 
Meissner state. The only qualitative  
difference with respect to the results of Ref.\cite{1} is that 
here we do not have lifetime effects.

	The spectral weights (Eqs. (\ref{poleslike})) are 
given by
\begin{eqnarray}\label{diaspecw}
\hat{\alpha}_1(\vec{k}) &=& \frac{[\omega_o(\vec{k}) - 
U(1-\rho_{\bar{\sigma}})][(\omega_o(\vec{k})+ \varepsilon_{{\vec{k}}})
(\omega_o(\vec{k}) + U) - \rho_{\bar{\sigma}}U \varepsilon_{{\vec{k}}}]}
{2\omega_o(\vec{k})(\omega_o^2(\vec{k}) - \omega^2_1(\vec{k}))}~~~,
\nonumber\\
\hat{\alpha}_2(\vec{k}) &=& \frac{[\omega_o(\vec{k}) + 
U(1-\rho_{\bar{\sigma}})][(\omega_o(\vec{k})- \varepsilon_{{\vec{k}}})
(\omega_o(\vec{k}) - U) - \rho_{\bar{\sigma}}U \varepsilon_{{\vec{k}}}]}
{2\omega_o(\vec{k})(\omega_o^2(\vec{k}) - \omega^2_1(\vec{k}))}~~~,
\nonumber\\
\hat{\alpha}_3(\vec{k}) &=& \frac{[\omega_1(\vec{k}) - 
U(1-\rho_{\bar{\sigma}})][(\omega_1(\vec{k})+ \varepsilon_{{\vec{k}}})
(\omega_1(\vec{k}) + U) - \rho_{\bar{\sigma}}U \varepsilon_{{\vec{k}}}]}
{2\omega_1(\vec{k})(\omega_1^2(\vec{k}) - \omega^2_o(\vec{k}))}~~~,
\nonumber\\
\hat{\alpha}_4(\vec{k}) &=& \frac{[\omega_1(\vec{k}) + 
U(1-\rho_{\bar{\sigma}})][(\omega_1(\vec{k})- \varepsilon_{{\vec{k}}})
(\omega_1(\vec{k}) - U) - \rho_{\bar{\sigma}}U \varepsilon_{{\vec{k}}}]}
{2\omega_1(\vec{k})(\omega_1^2(\vec{k}) - \omega^2_o(\vec{k}))}~~~.
\end{eqnarray}
\noindent and

\begin{eqnarray}\label{offspecw}
\hat{\beta}_1(\vec{k}) &=& \frac{\alpha(T)[\omega_o(\vec{k}) -
U(1-\rho_{\bar{\sigma}})]}
{2\omega_o(\vec{k})(\omega_o^2(\vec{k}) - \omega^2_1(\vec{k}))} 
\equiv \alpha(T) \bar{\beta}_1(\vec{k})~~~,
\nonumber\\
\hat{\beta}_2(\vec{k}) &=& \frac{\alpha(T) [\omega_o(\vec{k}) +
U(1-\rho_{\bar{\sigma}})]}
{2\omega_o(\vec{k})(\omega_o^2(\vec{k}) - \omega^2_1(\vec{k}))} 
\equiv \alpha(T) \bar{\beta}_2(\vec{k})~~~,
\nonumber\\
\hat{\beta}_3(\vec{k}) &=& \frac{\alpha(T)[\omega_1(\vec{k}) -
U(1-\rho_{\bar{\sigma}})]}
{2\omega_1(\vec{k})(\omega_1^2(\vec{k}) - \omega^2_o(\vec{k}))} 
\equiv \alpha(T) \bar{\beta}_3(\vec{k})~~~,
\nonumber\\
\hat{\beta}_4(\vec{k}) &=& \frac{\alpha(T)[\omega_1(\vec{k}) +
U(1-\rho_{\bar{\sigma}})]}
{2\omega_1(\vec{k})(\omega_1^2(\vec{k}) - \omega^2_o(\vec{k}))} 
\equiv \alpha(T) \bar{\beta}_4(\vec{k})~~~.
\end{eqnarray}

	Now it is an easy matter to convince ourselves that the 
following relations are satisfied
\begin{eqnarray}\label{sumrules}
\hat{\alpha}_1(\vec{k}) + \hat{\alpha}_2(\vec{k}) + \hat{\alpha}_3(\vec{k}) 
+ \hat{\alpha}_4(\vec{k}) &=& 1~~~,\nonumber\\
\omega_o(\vec{k})(\hat{\alpha}_1(\vec{k})-\hat{\alpha}_2(\vec{k})) + 
\omega_1(\vec{k})(\hat{\alpha}_3(\vec{k})-\hat{\alpha}_4(\vec{k})) &=& 
\varepsilon_{{\vec{k}}} + \rho_{\bar{\sigma}}U~~~,\nonumber \\
\hat{\beta}_1(\vec{k}) + \hat{\beta}_2(\vec{k}) + 
\hat{\beta}_3(\vec{k}) + \hat{\beta}_4(\vec{k}) &=& 0~~~,
\end{eqnarray}
\noindent
which are the first three sum rules for the moments\cite{1,Nolting}. The 
second off-diagonal moment is not satisfied because our order 
parameter is not $\Delta(T)$ but $\alpha(T)$. A discussion of 
the {\it failure} of a Hubbard I-type solution (for $\alpha(T) = 0$), 
as the one presented in Eqs. (\ref{solutions}), has 
been pointed out by Laura Roth\cite{Laura} many years ago, where 
she remarked that the diagonal Hubbard I solution only 
satisfied the first two sum rules for the moments. We 
should point out that the moment solution and the Hubbard I-type 
solution are different approaches, which turn out to be 
approximately equal due to the fact that some of the first 
sum rules are satisfied. Laura Roth's criticism to the Hubbard--I 
solution is still valid in the present calculation since we should 
have a correlation gap for any {\it finite} value of $U/t$. In order 
to close this gap, lifetime effects are called for.

 	From the spectral theorems we have\cite{Mahan}
\begin{equation}\label{spectral}
\left< c_{i \sigma} c^{\dagger}_{j \sigma} \right> =
\int_{-\infty}^{+\infty} \frac{A_{i,j}(\omega) d \omega}
{exp(\beta \omega) + 1}~~~;
~~~\left< c_{i \uparrow} c_{j \downarrow}\right> =
\int_{-\infty}^{+\infty}
\frac{B_{i,j}(\omega) d \omega} {exp(\beta \omega) + 1}~~~.
\end{equation}

	Then, we have
\begin{equation}\label{510}
\rho = \frac{1}{N_s}\sum_{\vec{k}}\int_{-\infty}^{+\infty} 
\frac{A({\vec{k}},\omega) d\omega}{exp(\beta \omega) + 1}~~~,~~~
\alpha(T) = \frac{1}{N_s}\sum_{\vec{k}}\int_{-\infty}^{+\infty} 
\frac{\varepsilon({\vec{k}})B({\vec{k}},\omega) d\omega}
{exp(\beta \omega) + 1}~~~,
\end{equation}
\noindent where $\beta = 1/T$ is the inverse of the temperature 
and $\alpha(T)$ has been defined just after Eq. (\ref{123123}).  
$A({\vec{k}},\omega)$ and $B({\vec{k}},\omega)$ are the 
one--particle spectral densities given as
\begin{eqnarray}\label{specdensities}
A({\bf k},\omega) \equiv - \frac{1}{\pi}
\lim_{\delta \rightarrow 0^+}Im[G({\bf k},\omega+i\delta)]~~; ~~~~~
B({\bf k},\omega) \equiv - \frac{1}{\pi}
\lim_{\delta \rightarrow 0^+}Im[F({\bf k},\omega + i\delta)]~~. ~~~
\end{eqnarray}
 
	In consequence, by combining Eqs. 
(\ref{spectral},\ref{510},\ref{specdensities}) 
with our Green functions (Eqs. (\ref{solutions})), 
we obtain the following self-consistent equations
\begin{eqnarray}\label{selfconseqs}
\frac{1}{U} &=& \int_{-4}^{+4}\varepsilon N(\varepsilon) 
\left[ \frac{\bar{\beta}_1(\varepsilon) + \bar{\beta}_2(\varepsilon)
exp(\beta \omega_o(\varepsilon))}{exp(\beta \omega_o(\varepsilon)) 
+ 1} + \frac{\bar{\beta}_3(\varepsilon) + \bar{\beta}_4(\varepsilon)
exp(\beta \omega_1(\varepsilon))}{exp(\beta \omega_1(\varepsilon)) 
+ 1} \right] d\varepsilon ~~~;\nonumber \\ 
\rho &=& \int_{-4}^{+4} N(\varepsilon)
\left[ \frac{\hat{\alpha}_1(\varepsilon) + \hat{\alpha}_2(\varepsilon)
exp(\beta \omega_o(\varepsilon))}{exp(\beta \omega_o(\varepsilon))
+ 1} + \frac{\hat{\alpha}_3(\varepsilon) + \hat{\alpha}_4(\varepsilon)
exp(\beta \omega_1(\varepsilon))}{exp(\beta \omega_1(\varepsilon))
+ 1} \right] d\varepsilon ~~~,
\end{eqnarray}
\noindent
where $N(\varepsilon)$ is the 2D density of states, $t = 1$, and 
we have chosen $ \rho = \rho_{\sigma} = \rho_{\bar{\sigma}}$, i.e., 
we are in the paramagnetic phase. 

     We indicate that our Eqs. 
(\ref{selfconseqs}) will respect particle-hole symmetry\cite{SRN}.   
From the analysis of the first equation of Eq. (\ref{selfconseqs}) 
and the definition of the our {\it order parameter} we can 
conclude that it allows any type of s--type of wave symmetry, 
since when performing the $\vec{k}$--integration of $\Delta(k)$ 
$\times$ $\epsilon(\vec{k})$, we see that $\alpha(T) \neq 0$ only 
if $\Delta(\vec{k})$ is of s--type. So, 
in this case, Eq. (\ref{selfconseqs}) can give rise to an order 
parameter of symmetry different from {\it pure} s-wave, a conclusion 
which was reached in a previous work\cite{RCD} 
using the sum rules both for the diagonal and the 
off--diagonal one--particle spectral functions. However, for 
a {\it pure s--wave}, i.e., $\Delta(\vec{k}) = const.$, we get 
$\alpha(T) \equiv 0$. So, our approximation fails\cite{disc} and we must 
go back to Eq. (\ref{approximation}) and directly perform our 
approximation in 
$\Gamma_2(\vec{k},\omega)$. We arrive to the mean--field 
$BCS$ results, where the chemical potential gets renormalized 
by the Hartree shift, i.e., $\rho U$. Due to our 
lazyness of languaje, we have used the word {\it order parameter} 
in this paragraph to denote $\Delta(\vec{k})$, even though 
we defined $\alpha(T)$ just after Eq. (\ref{offcoupling}) 
as the order parameter of our theory. Naturally, $\alpha(T)$ 
is the integral of $\Delta(\vec{k})$ weighted with $\varepsilon(\vec{k})$. 
So, they are related in some way, i.e., if $\Delta(\vec{k}) \equiv 
0$, then $\alpha(T) = 0$.

	In short, using the Hubbard--I decoupling scheme for 
both the diagonal and off-diagonal one-particle Green functions 
we have shown that these Green functions have four poles, 
symmetric in pairs, which qualitatively {\it verify} the more 
elaborated calculation of Ref.\cite{1}, as it has been previously 
discussed. Our one--particle Green functions satisfy sum rules 
for the moments and we have obtained other symmetries than a  
{\it pure}  s--wave order parameter. The range of validity 
of our approximation is contained in Eqs. (\ref{DysonGamma},
\ref{approximation}). Now we solve our Eqs. (\ref{selfconseqs}) 
in a low order approximation: We fix the value of our 
{\it order parameter}, $\alpha(T)$, and find the chemical 
potential using the first of Eqs. (\ref{selfconseqs}). For 
$\rho = 0.01$, $U/t = -8.0$ and $\frac{\alpha(T)}{t^2} = 0.1$, 
we find $\mu/t \approx -2.825$. We should say that our approach 
is valid for $|U| \geq W$, where $W = 8t$ is the bandwidth in 
two dimensions. Our calculation have been 
performed for $\beta = 1/T = 100.0$. We leave for the future\cite{JJRNpre}  
the numerical evaluation of the critical temperature, $T_c$, and 
the order parameter as function of temperature for different 
values of $U/t$ and electron concentration. 

	In order to deal 
with $d$--wave superconductivity, we should study a nearest 
neighboor ($n.n.$) attractive interaction. This model has been 
considered few years ago, at the mean field level, by 
Meintrup, Schneider and Beck\cite{Thomas} (See, also Ref.\cite{Chen}). 
We mention, while 
leaving, that lifetime effects can be included in a natural 
way both in $G(\vec{k},\omega)$ and $F(\vec{k},\omega)$, as it 
has been previously done in Ref.\cite{JJSS} for the self--energy. 
Work along these lines is in progress.
\begin{center}
{\large Acknowlegments}
\end{center}

    We would like to thank CNPq--Brazil  (project No. 
300705/95-6), FAPERGS--Brazil, the Swiss National Science Foundation,   
and  CONICIT--Venezuela (project F-139).  
Prof. Mucio Continentino indicated us that the study of this 
problem would be a rewarding task.  
We thank Mar\'{\i}a Dolores Garc\'{\i}a
for reading  the manuscript.\\
%
%
%
%


\begin{references}
\bibitem{Bednorz-Muller}
      	J. Bednorz and K.A. M\"uller,
      	Z. Phys. B~{\bf 64}, 189 (1986)
\bibitem{Anderson}
	P.W. Anderson, Science {\bf 235}, 1196 (1987); 
	{\it Frontiers and Borderlines in Many Particle Physics}. 
	North Holland, Amsterdam (1988); ibid., {\it The 
	Theory of Superconductivity in the High--$T_c$ Cuprates}. 
	Princeton Series in Physics (1997); J. Gonz\'alez, 
	M.A. Mart\'{\i}n--Delgado, G. Sierra and A.H. Vozmediano, 
	{\it Quantum Electron Liquids and High--$T_c$ 
	Superconductivity}. Lecture Notes in Physics m 38. 
	Springer--Verlag (1995)
\bibitem{HubbardI}
 	J. Hubbard, Proc. R. Soc. London A {\bf 276}, 238 (1963)
\bibitem{HubbardII}
	J. Hubbard, Proc. R. Soc. London A {\bf 281}, 401 (1964)
\bibitem{Fulde}
	Peter Fulde, {\it Electron Correlations in Molecules 
	and Solids}. Springer-Verlag (1993). 2nd Edition
\bibitem{tasaki}
	Hal Tasaki, {\it The Hubbard Model: Introduction and 
	Some Rigorous Results}. Preprint Sissa (1995); ibid., J.
	Statistical  
	Phys. {\bf 84}, 535 9(1996)
\bibitem{miyakawa}
	N. Miyakawa, J.F. Zasadzinski, L. Ozyuzer, P. 
	Guptasarma, D.G. Hinks, C. Kendziora and K.E. Gray, 
	cond-mat/9809398
\bibitem{RandVar}
	M. Randeria, cond--mat/9710223; M. Randeria and J.-C. 
	Campuzano, cond--mat/9709107, Varenna Lectures; M. 
	Randeria, "Photoemission Spectroscopy", in {\it X 
	Trieste Workshop on Open Problems in Strongly 
	Correlated Electron Systems}. ICTP--Italy, July 20--31, 
	1998
\bibitem{Micnas}
        R. Micnas, J. Ranninger and S. Robaszkiewicz, Rev. Mod.
        Phys. {\bf 62}, 113 (1990); A. Alexandrov, J. Ranninger and
        S. Robaszkiewicz, Phys. Rev. B {\bf 33}, 4526 (1986);
        M. Rice and L. Sneddon, Phys. Rev. Lett. {\bf 47},
        689 (1981). The last two references were the
        first ones to use this model for the
        study of the bismuthate superconductors. See also, P.H.J.
        Denteneer et al, Europhys. Lett. {\bf 16}, 5 (1991).
        The last authors compare the understanding of correlations
        in the attractive Hubbard model to the role played in
        transitions and critical phenomena by the
        Ising model
\bibitem{Randeria}
	M. Randeria in {\it Bose Einstein Condensation}, A. 
	Griffin et al (Eds.). Cambridge University Press (1994)
\bibitem{HS}
	C. Huscroft and R.T. Scalettar, cond--mat/9606038
\bibitem{FW}
        A.L. Fetter and J.D. Walecka, {\it Quantum Theory of
        Many-Particle Systems}. McGraw-Hill, New York (1971)
\bibitem{1}
        M.H. Pedersen, J.J. Rodr\'{\i}guez--N\'u\~nez,
        H. Beck, T. Schneider and
        S. Schafroth. Z. Phys. B {\bf 103}, 21 (1997). However, 
	the $BCS$ symmetry is broken if we include {\it double 
	fluctuations} in the $T$--matrix formulation, as it has 
	been done by S. Schafroth and J.J. Rodr\'{\i}guez--N\'u\~nez, 
	Z. Phys. B {\bf 102}, 493 (1997)
\bibitem{Roland}
	R. Kirchhofer, Dipl\^ome. Universit\'e de Neuch\^atel 
	(1997, unpublished); R. Kirchhofer, R. Fr\'esard, 
	H. Beck and J.J. Rodr\'{\i}guez--N\'u\~nez, Physica B 
	(1998, accepted), SCES'98; ibid., unpublished
\bibitem{EA}
	E.V. Anda, "The Metal Insulator Transition in the 
	Hubbard Model". {\it The Physics and  Mathematical Physics
        of the Hubbard Hamiltonian}. Ed. D. Campbell and F. Guinea. 
        Plenum Press, 1994
\bibitem{JJRNpre}
	J.J. Rodr\'{\i}guez--N\'u\~nez, in preparation
\bibitem{Nolting}
        W. Nolting,
        Z. Physik {\bf 255}, 25 (1972); W. Nolting,
        {\it Grundkurs: Theoretische
        Physik. 7 Viel-Teilchen-Theorie.} Verlag Zimmermann-Neufang
        (Ulmen--1992); T. Hermann and W. Nolting, cond--mat/9702022 
	and J. Magn. Magn. Mater. {\bf 170}, 253 (1997); 
        T. Schneider, M.H. Pedersen and
        J.J. Rodr\'{\i}guez--N\'u\~nez, Z. Phys. B {\bf 100}, 263 (1996)
\bibitem{Laura}
	Laura Roth, Phys. Rev. {\bf 184}, 451 (1969)
\bibitem{Mahan}
	G.D. Mahan, {\it Many--Particle Physics}. Second Edition. 
	Plenum (1990)
\bibitem{SRN}
	M.P. S{\o}rensen and J.J. Rodr\'{\i}guez--N\'u\~nez, 
	Physica C {\bf 274}, 323 (1997). The authors discuss 
	the $t-J$ model within the moment approach
\bibitem{RCD}
	J.J. Rodr\'{\i}guez--N\'u\~nez, C.E. Cordeiro and 
	A. Delfino, Physica A {\bf 232}, 408 (1996). The 
	results of this paper are in agreement with the 
	ones of the double fluctuation calculations of 
	Ref.\cite{1}
\bibitem{disc}
	Due to our approximation (Eq. (\ref{approximation})), 
	$\Delta(\vec{k}) = const.$ $\Longrightarrow$ 
	$\alpha(T) = 0$. There is the possibility of still 
	going beyond a mean--field treatment if 
	Eq. (\ref{approximation}) is not satisfied
\bibitem{Thomas}
	Th. Meintrup, Ph. D. thesis, Universit\'e de 
	Neuch\^atel, 1995 (unpublished); Th. Meintrup,  
	T. Schneider and H. Beck, Europhys. Lett. {\bf 31}, 231 (1995)
\bibitem{Chen}
	Q. Chen, I. Kosztin, B. Jank\'o and K. Levin, 
	cond--mat/9807414. They have considered the 
	"pairing approximation" to explain some 
	superconducting properties in underdoped to 
	overdoped cuprates
\bibitem{JJSS}
	J.J. Rodr\'{\i}guez--N\'u\~nez and S. Schafroth, 
	J. Phys.: Condens. Matter {\bf 10}, L391 (1998); 
	S. Schafroth and J.J. Rodr\'{\i}guez--N\'u\~nez, 
	to be submitted;  
        J.J. Rodr\'{\i}guez--N\'u\~nez, in preparation	
\end{references}
\end{document}